# Hydrogen Diffusion and Stabilization in Single-crystal VO$_2$ Micro/nanobeams by Direct Atomic Hydrogenation

Jian Lin,[†,‡#] Heng Ji,[§#] Michael W. Swift,[§#] Will J. Hardy,[§] Zhiwei Peng,[⊥] Xiujun Fan,[‡,⊥] Andriy H. Nevidomskyy[§], James M. Tour[†,‡,⊥] and Douglas Natelson[§,‡,¶*]

[†]*Department of Mechanical Engineering and Material Science;* [‡]*Smalley Institute for Nanoscale Sci. & Tech.;* [§]*Department of Physics and Astronomy;* [⊥]*Department of Chemistry;* [¶]*Department of Electrical and Computer Engineering, Rice University, 6100 Main St., Houston, Texas 77005*

\# *These people contributed equally to this work.*

*\*To whom correspondence should be addressed. Emails:* natelson@rice.edu

**ABSTRACT**

We report measurements of the diffusion of atomic hydrogen in single crystalline VO$_2$ micro/nanobeams by direct exposure to atomic hydrogen, without catalyst. The atomic hydrogen is generated by a hot filament, and the doping process takes place at moderate temperature (373 K). Undoped VO$_2$ has a metal-to-insulator phase transition at ~340 K between a high-temperature, rutile, metallic phase and a low-temperature, monoclinic, insulating phase with a resistance exhibiting a semiconductor-like temperature dependence. Atomic hydrogenation results in stabilization of the metallic phase of VO$_2$ micro/nanobeam down to 2 K, the lowest point we could reach in our measurement setup. Optical characterization shows that hydrogen atoms prefer to diffuse along the *c*-axis of rutile (*a*-axis of monoclinic) VO$_2$, along the oxygen "channels". Based on observing the movement of the hydrogen diffusion front in single crystalline VO$_2$ beams, we estimate the diffusion constant for hydrogen along the *c*-axis of the rutile phase to be $6.7 \times 10^{-10}$ cm$^2$/s at approximately 373 K, exceeding the value in isostructural TiO$_2$ by ~ 38×. Moreover, we find that the diffusion constant along *c*-axis of the rutile phase exceeds that along the equivalent *a*-axis of the monoclinic phase by at least three orders of magnitude. This remarkable change in kinetics must originate from the distortion of the "channels" when the unit cell doubles along this direction upon cooling into the monoclinic structure. *Ab initio* calculation results are in good agreement with the experimental trends in the relative kinetics of the two phases. This raises the possibility of a switchable membrane for hydrogen transport.

Vanadium dioxide is a strongly correlated material that undergoes a first-order metal-insulator transition (MIT) from a high temperature, rutile, metallic phase to a monoclinic, insulating phase as the temperature goes below 340 K[1-4]. There has been considerable debate over the underlying mechanism of the transition, including the relative contributions of Peierls-like[5] (lattice distortion due to electron-phonon interaction) and Mott-like[2] (strong electron-electron correlation) physics.[5-8] In addition, this material has attracted much attention for its potential applications in ultrafast optical and electrical switching.[9-13] Therefore, much effort has been made to modulate the MIT by various approaches such as strain engineering,[14, 15] ionic liquid gating,[16-18] electric field-induced oxygen vacancy,[19] and chemical doping.[20-25]

In terms of chemical doping, using metal elements such as tungsten during growth can dramatically change the transition properties[20, 26], but it is an irreversible process. Recently, we have demonstrated that reversible doping with atomic hydrogen alters the electronic phase transition.[18, 24] Further work has reported atomic hydrogen doping via high temperature annealing,[22, 23, 25, 27] confirming that oxygen depletion[19, 27] is not taking place in these cases, and determined the stable structure of the hydrogenated material.[27] Despite this progress, the need for atomic rather than molecular hydrogen required either high temperature annealing[22, 23, 25] or Pd catalysts,[24, 27] which either complicates the mechanism by inducing possible side chemical reactions, or limits practical applications. For example, in Refs. 24 and 27 the temperature of hydrogenation using the spillover method with Pd catalyst was > 100 °C which is above phase transition temperature ($T_t$).[24] It remains unclear whether the elevated temperature is required to enhance the kinetics of the catalyst, to thermally enhance the diffusivity of the resulting atomic hydrogen, or whether the rutile crystal structure is of critical importance to the rate of the diffusion process.   To our knowledge there have also been no quantitative studies of the diffusion kinetics of hydrogen in $VO_2$, in either the high temperature rutile or low temperature monoclinic phases.   Here we report measurements of the kinetics of hydrogenation of $VO_2$ in the absence of any catalyst, eliminating unknown catalytic kinetics, possible conflicting reactions, and allowing hydrogen access to all crystallographic surfaces.  We compare the hydrogen diffusion rate in monoclinic and rutile phases of single crystalline $VO_2$ micro/nanobeams, based on optical detection of metallic domains.

Consistent with prior work involving catalytic spillover, we observe hydrogen diffusion and stabilization of the metallic state in single crystalline $VO_2$ micro/nanobeams by direct atomic hydrogen generated by a hot filament while maintaining low sample temperatures (~100 °C). The atomic hydrogenation results in stabilization of the metallic phase of $VO_2$ micro/nanobeams down to the lowest measured temperature (~ 2 K).   Direct imaging of the nanobeams allows us, through the difference in optical properties between metallic and insulating states, to measure the rate of rapid hydrogen diffusion along the beam growth direction (rutile *c*-axis; monoclinic *a*-axis), with no detectable evidence for hydrogen diffusion in the transverse directions.  In analogy with rutile $TiO_2$, this anisotropy of diffusion is expected[28, 29], as we discuss below.  We find that the diffusion rate along the favored direction depends very strongly on whether the material is in the monoclinic or rutile phases. The rate of migration of the H-stabilized metallic phase boundary indicates that the diffusion constant *D* is at least three orders of magnitude larger in the rutile phase than that in monoclinic phase, raising the possibility of VO$_2$ as a switchable semipermeable membrane for hydrogen transport. *Ab initio* calculations of the barrier for hydrogen diffusion along this favored direction are consistent with the experimental data's contrast between the rates in the monoclinic and rutile states.

The micro/nanobeams were grown on silicon wafers with 2 μm thermal oxide via physical vapor deposition (PVD), as described elsewhere.[30] The detailed synthesis protocol can be found in the experimental method. Pure VO$_2$ micro/nanobeams with rectangular cross sections nucleate on the wafer surface and grow in their rutile [001] direction (which becomes the monoclinic [100] when the beam is cooled below the transition temperature).[30] Loosened from the growth substrate by buffered oxide etching (mixture of 40% NH$_4$F solution and 49% HF solution with volume ratio 6:1) and subsequently transferred to a carrier Si/SiO$_2$ substrate to minimize strain, the beams exhibited the insulating state with monoclinic structure below the MIT temperature of ~340 K (Figure 1a). When the temperature was increased above the transition temperature (345 K) the beams turned into the fully metallic state with rutile structure,[24] as inferred from the color change in Figure 1b. This same sample was hydrogenated by direct exposure to atomic hydrogen generated by a hot filament for a certain period, with the sample temperature maintained at 373±10 K, followed by cooling to room temperature in the absence of hydrogen (see the schematic in Figure S1 and the methods section for the experimental setup and details). Figure 1c shows distinct regions of darker color on both ends of the VO$_2$ nanobeam after 5 min of exposure to atomic hydrogen at a sample temperature elevated above the phase transition temperature. These regions are in a hydrogen-stabilized metallic state, as observed through their differing optical contrast and Raman spectra.  This partial conversion of the beams to a hydrogen-stabilized metallic state agrees well with the results seen when catalytic spillover is used to introduce hydrogen at the ends of similarly prepared nanobeams.[24]  In contrast to the spillover situation, in this case atomic hydrogen had access to all exposed faces of the beam.  Based on optical signatures of the beam state, hydrogen diffusion in directions transverse to the beam growth direction (i.e., normal to the beam sidewalls) is clearly far slower than diffusion along the beam growth direction from the exposed ends of the beam.  Such anisotropy in diffusion is reminiscent of the diffusion of atomic hydrogen in the isostructural rutile titanium oxide[28]. After the $VO_2$ nanobeams were hydrogenated under identical substrate temperatures and gas flow for times approaching 15 min, the dark blue regions on both ends extended toward each other and fully occupied the beams (Figure 1d).

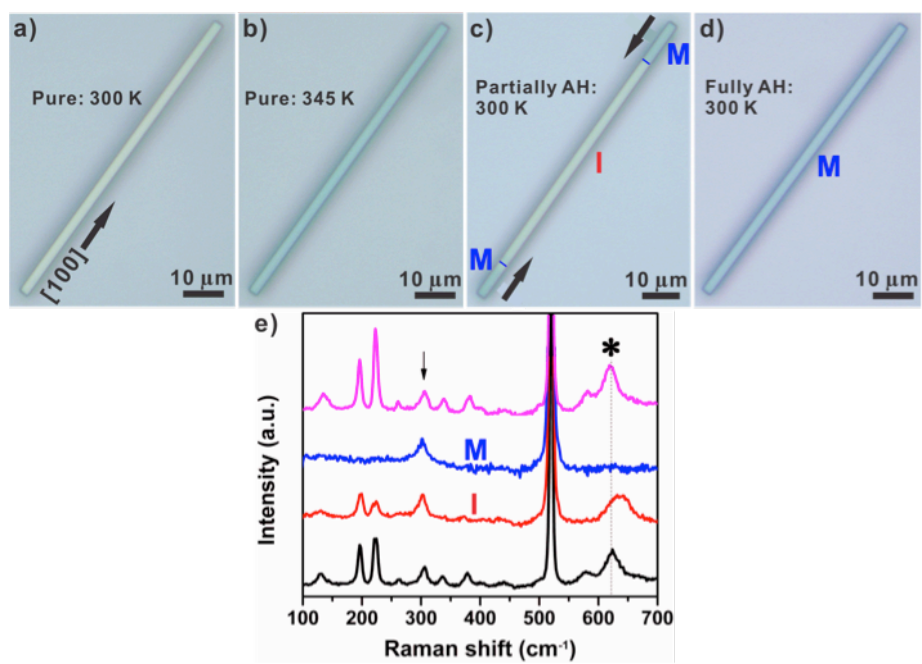

**Figure 1. Optical and Raman spectra of $VO_2$ nanobeams.** (a) Optical image of pure $VO_2$ taken at 300 K. (b) Optical image of pure $VO_2$ taken at 345 K. (c) Optical image of $VO_2$ taken at 300 K after hydrogenation at 100±10 °C for 5 min.  Lines have been drawn to indicate clearly the boundaries between metallic domains (propagating in from the ends) and the insulating central region.   (d) Optical image of $VO_2$ taken at 300 K after hydrogenation at 100±10 ºC for 15 min. e) Raman spectra of pure $VO_2$ crystals (black); of the light-colored regions of the hydrogenated $VO_2$ (labeled as "I" in **c**, red); of the dark-colored regions of the hydrogenated $VO_2$ (labeled as "M" in **c** and **d**, blue); of hydrogenated $VO_2$ after baking in air at 250 ºC for 20 min (magenta). All of the spectra were taken at 300 K.

Raman spectroscopy was used to investigate the structural phases of the pure and hydrogenated $VO_2$ crystals. The phases can be identified by monitoring the peaks in the black spectrum (Figure 1e).   This Raman spectrum of the as-grown $VO_2$ exhibits distinct peaks at 196, 222, 262, 337, 379 and 623 cm$^{-1}$ with additional peaks due to the silicon substrate at 302 and 521 cm$^{-1}$, indicating the stable monoclinic phase (M1) at room temperature and in good agreement with previous reports[24, 31]. The full width at half-maximum (FWHM) of the particular peak (indicated by an asterisk) at 623 cm$^{-1}$ is 23.4 cm$^{-1}$. When the $VO_2$ nanobeams were partially hydrogenated as shown in Figure 1c, the Raman spectra from the distinct regions show different characteristics. The Raman spectrum taken from the light colored central region of the beam (red curve in Figure 1e) is similar to that of the M1 phase. However, the indicated peak shifted to 637 cm$^{-1}$ and its FWHM increased to 46.8 cm$^{-1}$. This may be an indication of the light doping in this region, such that the concentration of hydrogen present is not large enough to stabilize the metallic state at room temperature. Raman spectra from the dark colored regions of the hydrogenated $VO_2$ nanobeams (blue curve in Figure 1e) are dominated by weak and broad bands, resembling spectra of $VO_2$ in the rutile phase (R)[24, 32]. Baking a previously fully hydrogenated $VO_2$ nanobeam in air at 250

°C for 20 min restores its Raman spectrum to be nearly identical to that of the as-grown monoclinic state, indicating excellent reversibility of the atomic hydrogenation.

To confirm the crystal structure of the hydrogenated $VO_2$, transmission electron microscopy (TEM) and selected area electron diffraction (SAED) patterns were performed on nanobeams, as shown in Figure S2. Figure S2a,b shows the TEM images of a pure $VO_2$ nanobeam in low and high magnification. The clear lattice fringe in Figure S2b indicates the single crystallinity of $VO_2$ nanobeams. The inset SAED pattern in Figure S2b agrees well with the monoclinic structure and longitudinal growth direction of the nanobeam along the [100] direction[30]. After hydrogenation, $VO_2$ nanobeams remain highly crystalline and show no visible non-uniformity (Figure S2c).

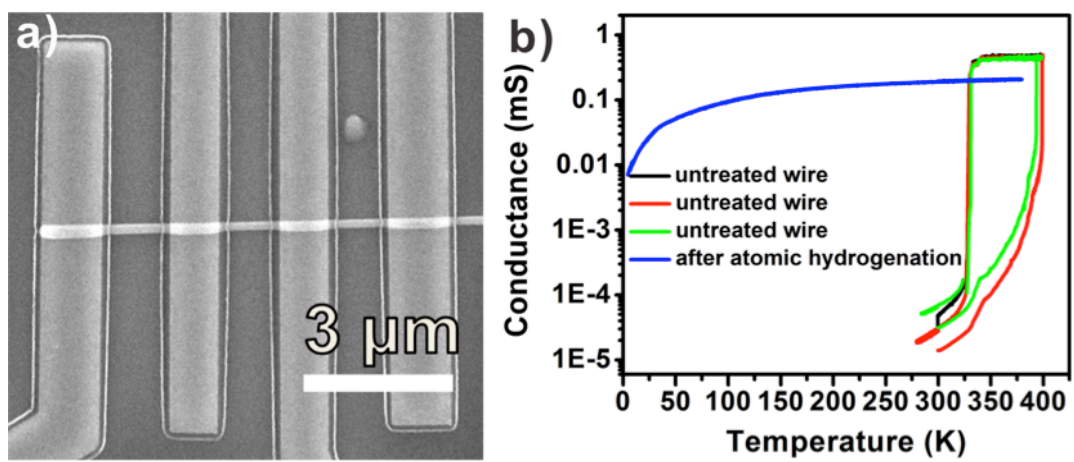

**Figure 2. Electronic transport of $VO_2$ nanobeam devices**. (a) Optical image of a four-terminal $VO_2$ nanobeam device. (b) Conductance (log scale) versus T for pure $VO_2$ nanobeams (black, red and green curves) and a typical one that was hydrogenated above the phase transition temperature for 15 min (blue curve).

To demonstrate the stabilization of metallic $VO_2$ nanobeams by atomic hydrogen, four-terminal electron devices were fabricated to study the temperature dependent electrical conductance. Figure 2a shows a representative device fabricated on a silicon substrate with 2 μm thermal oxide. Before hydrogenation, the conductance of the $VO_2$ devices changes by a factor of over $10^3$ at the metal-insulator transition temperature of ~ 395 K and shows the hysteresis during the cooling and heating cycles. This increased transition temperature compared with bulk material at 341 K is due to the built-in compressive strain in $VO_2$ nanobeams embedded on substrates[14, 15, 24]. After being hydrogenated above the MIT temperature for 15 min, the phase transition vanishes and the devices show comparatively high conductivity down to 5 K, indicating relatively metallic character of hydrogenated $VO_2$. We note that unlike a traditional "good" metal state, the conductance versus temperature curve does not show a negative slope at low temperatures. Furthermore, as reported previously, there is device-to-device variability in conductance as a function of temperature, likely due to the residual strain of particular contacted beams.

We now consider the effect of substrate temperature on hydrogen diffusion. Figure S3a shows pure $VO_2$ nanobeams lying on and strongly mechanically coupled to their growth substrate. Strain-induced metal-insulator domains when heating the materials above 340 K can be distinguished, as reported previously (Figure S3b)[15]. Domain structures persist up to 380 K, above which the material is fully in the metallic state (Figure S3c). If the same samples were hydrogenated above phase transition temperature (380 K) for 20 min, the phase transition disappears and the metallic state is stabilized when cooling down to room temperature (Figure S3d-e). On the other hand, the room temperature hydrogenation process is much slower. The same $VO_2$ beams have been reused after baking in air at 250 °C for 30 min to retrieve the phase transition. Following hydrogenation at substrate temperatures below the phase transition temperature (*e.g.*, 300 K) even for 3 h, the nanobeams still have no visible boundary or color change (at least under 100 nm resolution) caused by hydrogenation, and when heated up through the transition, they exhibit phase domain patterns similar to those seen in the unexposed pure beams (Figure S4). The Raman spectrum also confirms that the $VO_2$ nanobeams exposed to atomic hydrogen below the phase transition temperature have the same characteristics as pure beams (Figure S5), suggesting that the beams remain in the monoclinic phase after attempted hydrogenation below the phase transition temperature.

Since these experiments were performed without any catalyst, the temperature-dependent kinetics observed here should reflect that of the $H-VO_2$ system. It is true that impurity diffusion in solids is generally thermally activated. However, the significant differences observed for the motion of hydrogen-stabilized phase boundaries when the diffusion process takes place in the rutile vs. the monoclinic state strongly suggests that the structural phase of $VO_2$ has an enormous impact on the rate of hydrogen diffusion. Empirically, hydrogen diffuses much faster along the *c*-axis of the rutile phase than along the (equivalent) *a*-axis of the monoclinic phase; attempts to diffuse H into beams in the monoclinic structure produce no observable hydrogen intercalation.

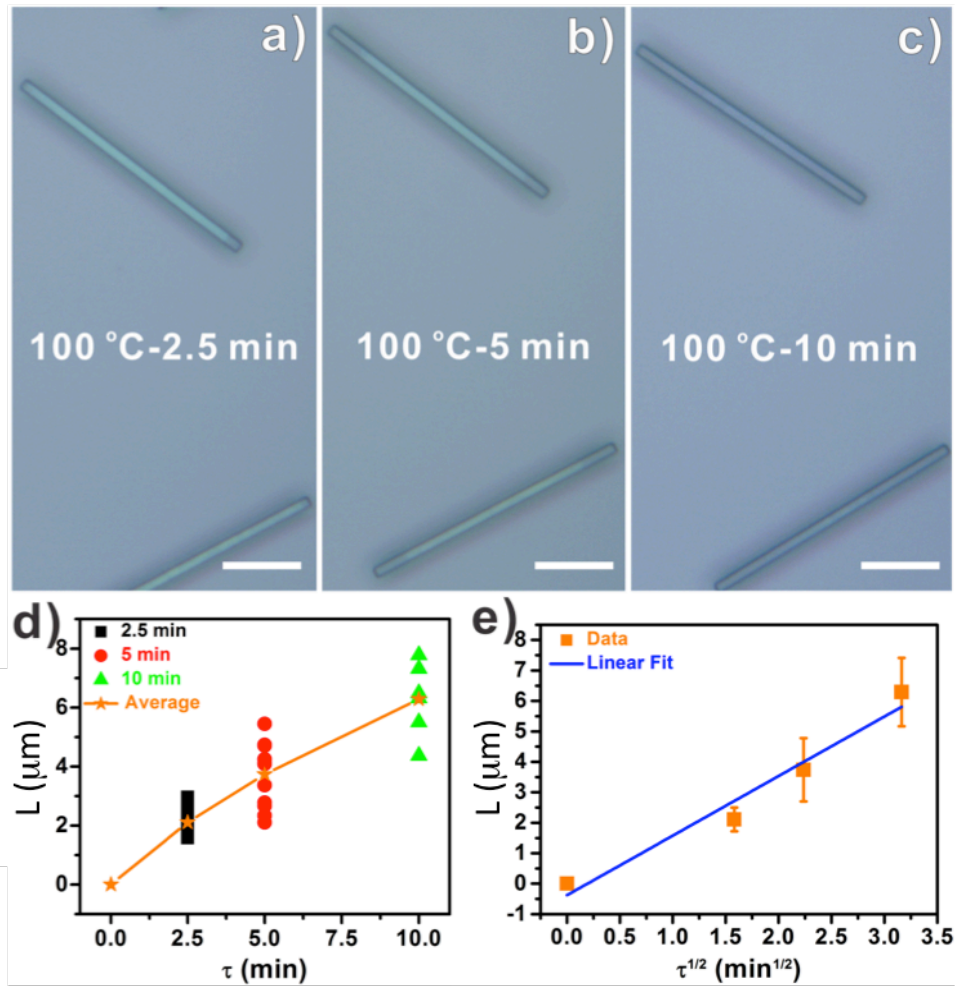

**Figure 3. Hydrogen diffusion in VO$_2$ beams.** (a-c) Representative optical images of a VO$_2$ beams following atomic hydrogenation at 100±10 ºC for 2.5 min, 5 min and 10 min; scale bar, 10 µm. (d) Plot of diffusion length (L$_{Diffusion}$) (estimated from the size of the stabilized metallic regions) versus hydrogenation time (τ) of rutile phase VO$_2$ beams. (e) Plot of average L$_{Diffusion}$ versus $\tau^{1/2}$ of rutile phase VO$_2$ beams.

Based on these observations, to quantify the diffusion process we systemically studied the hydrogen diffusion rates of VO$_2$ nanobeams by exposing the nanobeams to atomic hydrogen while in the monoclinic and rutile phases for various durations. Experiments to estimate diffusion kinetics were restricted to beams that did not show large effects of interfacial strain (e.g., breaking into metal/insulator domains upon thermal cycling as in Fig.

S3).  Figure 3a-c show the nanobeams (transferred to a carrier substrate, and thus less strained than those in Figure S3) exposed to atomic hydrogen while maintained in the rutile phase for 2.5, 5 and 10 min, respectively. As in Fig. 1, metallic domains indicated by darker color form at both ends and expand toward the beam centers as the exposure duration increases. Figure 3d shows the distance traversed by the boundary between the H-stabilized metallic domains and the monoclinic state, following various durations of exposure to atomic hydrogen while in the rutile state.  Figure 3e shows that the average diffusion length is approximately linearly proportional to square root of time, as expected for diffusion, where a typical diffusion path distance scale is $L = (D\tau)^{1/2}$ after a time $\tau$, and $D$ is the diffusion constant.

Assuming an effective 1d diffusion problem and that the domain boundary is a proxy for reaching a critical level of hydrogen concentration within the material, the slope of the linear fit in Fig. 3e allows us to determine the diffusion constant for H along the $c$-axis in rutile $VO_2$. With an uncertainty of a few percent from the fit, we find $D \approx 6.7 \times 10^{-10}$ cm$^2$/s. For comparison, extrapolating the expression of Johnson et al.[29] for the diffusion constant for H along the $c$-axis of isostructural rutile $TiO_2$ to 373 K, we find around $1.8 \times 10^{-11}$ cm$^2$/s. Thus diffusion of H along $c$ in rutile $VO_2$ is around 38× more facile than in rutile $TiO_2$. Note that Johnson et al. report for $TiO_2$ that the H diffusion constant transverse to the $c$ direction at 373 K is lower by a factor of ~ $10^7$ (!).   Thus it is not surprising that we observe no evidence of H diffusion transverse to the rutile $c$-axis in $VO_2$.

The motion of the boundary of the H-stabilized metallic region (on the order of microns per minute when hydrogenation takes place in the metallic state; less than 100 nm in three hours when hydrogenation takes place in the monoclinic state) suggests that the diffusion constant must be orders of magnitude larger when hydrogenation takes place with the beam in the rutile state than when the beam is exposed in the monoclinic state (Figure S4c-d). An alternate possibility is that the solubility of hydrogen in the monoclinic state is so poor that it is never possible for the monoclinic phase to accommodate enough hydrogen to trigger the transformation into the stabilized metallic structure. It is not clear that this is consistent with measurements of the phase diagram of the H-VO$_2$ system, however.[27] We also note that there is large beam-to-beam variability of the length of these domains (seen as the vertical spread in the data points in Figure 3d); we believe that this could be a manifestation of different strain conditions in the various beams modulating diffusion rates.

Overall, the trends are clear: (1) Atomic hydrogen diffuses rapidly (even more so than in the well-studied TiO$_2$ analog) along the rutile *c*-direction, with no detectable diffusion transverse to the rutile *c*-direction; (2) The diffusion constant for atomic hydrogen along the equivalent monoclinic *a*-axis is at least 1000× smaller than in the rutile structure, even though the temperature difference between the rutile and monoclinic experiments was only tens of K.

To further understand the physical mechanism of the difference in hydrogen diffusion rates between two phases, we performed *ab initio* structural calculations to estimate the energy scales associated with hydrogen diffusion. We modeled the intercalation of atomic hydrogen into the atomic lattice of VO$_2$ as a diffusion process. In solids, the diffusion constant *D* is well modeled by the Arrhenius equation[28, 33, 34].

$$D = D_0 e^{-E_{\text{diff}}/k_B T} \quad (1)$$

where $D_0$ is the infinite-temperature diffusivity and $E_{\text{diff}}$ is the energy barrier associated with diffusion of the impurity from site to site. Equivalently, the hopping frequency $\Gamma$ is given by

$$\Gamma = \frac{D}{d^2} = \nu e^{-E_{\text{diff}}/k_B T} \qquad (2)$$

where $d$ is the average jump distance and the $\nu = D_0/d^2$ is the attempt frequency[34].

Calculation of $D_0$ or $\nu$ would require detailed knowledge of the conditions outside the $VO_2$ lattice and an investigation of the vibrational modes of hydrogen ions within the lattice. We also note that full finite temperature modeling of the diffusion process would require a molecular dynamics component, keeping track of vibrational fluctuations in the vanadium and oxygen atomic positions. Such calculations are beyond the scope of this work. However, we note a recent molecular dynamics calculation of the diffusion barrier in the rutile phase[35]. Instead, we consider the ratio of the diffusion rates between rutile and monoclinic phases:

$$\frac{D(rut)}{D(mon)} = \frac{D_{0,rut}}{D_{0,mon}} e^{(E_{diff,mon} - E_{diff,rut})/k_B T} = \frac{D_{0,rut}}{D_{0,mon}} e^{\Delta E / k_B T} \qquad (3)$$

This value may then be compared with experiment. As we did in the analysis of Fig. 3e, we assume an essentially 1d diffusion problem along the crystallographic direction of interest, and that the metallic phase is stabilized when the local hydrogen concentration surpasses some threshold value. We can then estimate the size of a stabilized metallic domain as $\sim (D\tau)^{1/2}$, where $D$ is the diffusion constant in the relevant crystallographic phase, and $\tau$ is the duration of the diffusion process.

We seek to estimate the diffusion energy barrier $E_{\text{diff}}$ for hydrogen ions in the two different phases of $VO_2$. Details of the calculation are discussed in the Methods section. We use BFGS optimization to find the lowest energy position of a hydrogen atom within a *fixed* $VO_2$ lattice.  This assumes that the presence of hydrogen within the lattice at relatively low concentrations is only a small perturbation on the lattice structure in both the monoclinic and rutile phases.  Given that, we then move the hydrogen in small steps to a periodically equivalent point further down the rutile *c*/monoclinic *a*-axis, computing the total energy of the system at each point.  Note that this is equivalent to assuming that hydrogen motion is sufficiently fast that no structural relaxation takes place.  The difference between the maximum and minimum energies is the energy barrier for this path. The hydrogen will tend to take the path with the lowest energy barrier. We note that the energy barrier along the true path is actually a saddle point in three dimensions: a maximum along the channel direction, but a minimum in the plane transverse to the channel direction.

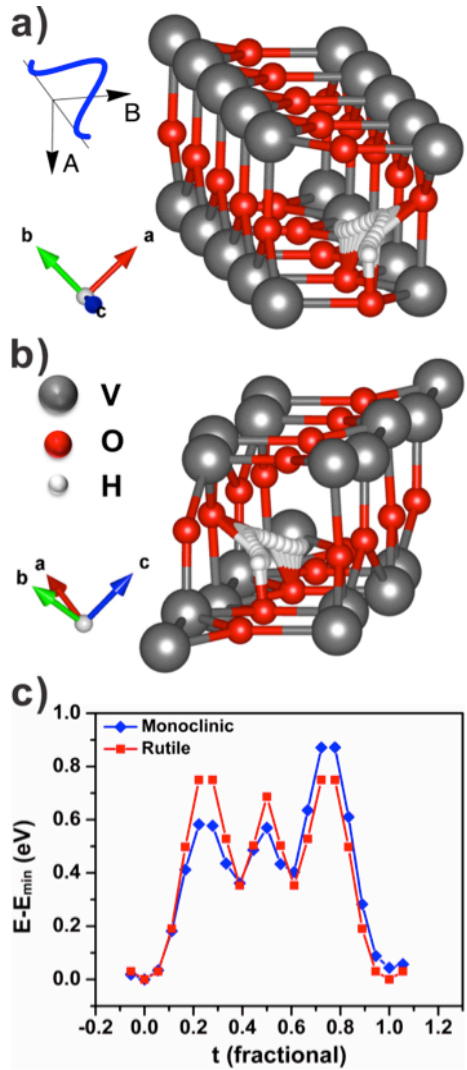

**Figure 4. *Ab initio* calculations**. **(a)** Lowest-energy path found for a hydrogen atom diffusing into rutile VO$_2$. The inset to subfigure (a) illustrates the sin$^2$ perturbation from a straight-line path scaled by parameters A and B. **(b)** Lowest-energy path found for a hydrogen atom diffusing into monoclinic VO$_2$. Several positions of the hydrogen atom at different points along the path are superimposed, showing the path stroboscopically. **(c)** The energy of the hydrogen as it travels through the channel along the optimized path. *t* is a parameter which shows the fractional progress along the path. Note that lattice positions (e.g, the vanadium sites) which had been equivalent in the rutile phase are no longer equivalent in the monoclinic phase because of the unit cell doubling.

We made an educated guess at the form of the path the hydrogen takes, which we then optimized. A straight-line path between two equilibrium hydrogen positions was perturbed with two orthogonal $\sin^2$ paths scaled by tunable parameters $A$ and $B$, which were optimized to give the minimum energy barrier (Figure 4a). The $\sin^2$ form was chosen based on the observation in the structure calculations that hydrogen in $VO_2$ tends to form an OH bond with the closest oxygen, so hydrogen will "swing" from oxygen to oxygen down the channel, with the swing amplitudes fixed by adjustable parameters $A$ and $B$. The $A$ direction is towards the oxygen bonded with hydrogen in the equilibrium (minimum energy) position, and the $B$ direction is orthogonal to the lattice, towards the closer wall of the channel. The magnitude is normalized to the length of the OH bond in the equilibrium position (1.02 Å). The path shown is the optimal one found for the rutile phase, with $A = -0.44$ and $B = 0.63$. The path for the monoclinic phase had $A = -0.25$ and $B = 0.60$.

The paths found are shown in Figure 4a and 4b. Figure 4c shows the energy of the system as the hydrogen is considered at various points through the channel along the optimized path. Here $t$ is a parameter which shows the fractional progress along the path. Note that lattice positions which had been equivalent in the rutile phase are no longer equivalent in the monoclinic phase. This is a result of the unit cell doubling in the monoclinic phase, and explains why the peak structure in Fig. 4c is symmetric about the middle of the path as a function of $t$ in the rutile phase, but asymmetric in the monoclinic phase. The numerical results for the diffusion barriers are:

$E_{\text{diff,mon}} = 0.8717 \pm 0.0027$ eV

$E_{\text{diff,rut}} = 0.7494 \pm 0.0053$ eV

$$\Delta E = 0.1223 \pm 0.0059 \text{ eV}$$

The activation energy for hydrogen diffusion in TiO$_2$ is ~ 0.6 eV[28], confirming that the results found here are reasonable. We will compare the diffusion rates at the transition temperature, T = 340 K. At this temperature, following Equation 3, we predict that the ratio of the diffusivities between the rutile and monoclinic phases should be $\frac{D(rut)}{D(mon)} = \frac{D_{0,rut}}{D_{0,mon}} e^{\Delta E/k_B T} = 65 \pm 13$, if the prefactors are assumed to be identical. These calculations show that, *on energetic grounds alone*, one should expect hydrogen diffusion to be much faster along this favored direction in the rutile state than in the monoclinic structure. We stress that our calculation provides an upper bound on the diffusion barrier. The fact that the experimental ratio inferred from the motion of the boundary of the hydrogen-stabilized metallic region is much larger suggests that there is additional physics at work beyond what has been included in the calculation, presumably enhancing diffusion in the rutile structure. A large difference in the *prefactors* seems unlikely, given that the attempt frequency for hydrogen hopping is likely set largely by the H..O interaction, which should be similar in the two structural phases. A likely candidate would be vibrational dynamics known to lead to greatly enhanced diffusion rates in TiO$_2$.[34] A recent *ab initio* molecular dynamics calculation[35] indeed finds the diffusion barrier in rutile VO$_2$ to be ~ 0.4 eV, significantly lower than the one found here.

In summary, the atomic hydrogen generated directly from a hot filament was used to stabilize the metallic phase of the VO$_2$ nanobeams. The phenomenon that hydrogen diffuses preferentially along the *a*-axis of monoclinic VO$_2$ (*c*-axis rutile VO$_2$) was observed, consistent with past studies of hydrogen motion in rutile TiO$_2$. The diffusion constant for hydrogen along this direction in rutile $VO_2$ is even larger than that in $TiO_2$. Notably, the propagation of the front of hydrogen-stabilized metallic phase along the *c*-axis of rutile phase is at least three orders of magnitude faster than along the equivalent *a*-axis of monoclinic $VO_2$. Zero-temperature theoretical calculations are consistent with the idea that the activation barrier for hydrogen diffusion is considerably lower along this direction in the rutile structure than in the monoclinic phase. However, the enormity of the difference in apparent diffusion kinetics between the two phases suggests that the zero-temperature calculations underestimate the diffusion rate in the rutile structure. The contrast in hydrogen diffusion rates between the two structural phases suggests that $VO_2$ could be used as a switchable semipermeable membrane for hydrogen transport. Moreover, hydrogenation of $VO_2$ through direct exposure to atomic hydrogen demonstrates another effective way of modulating the electrical, chemical and optical properties of the $VO_2$, potentially paving the way for $VO_2$ devices with new functionalities.

**METHODS**

**Synthesis and atomic hydrogenation of single-crystal VO$_2$ micro- and nanometer size beams.** VO$_2$ beams were synthesized by physical vapor deposition.[30] V$_2$O$_5$ powder was put in the ceramic boat loaded in the center of the heating zone of the furnace. The silicon substrate with 2 μm thermal oxide was located 3-4 cm away from the boat in the downstream flow. The nanobeams were grown under low vacuum. While flowing Ar carrier gas, the furnace temperature was ramped to 900 ºC, then maintained for 30 min followed by cooling to room temperature. Direct atomic hydrogenation was carried out in a vacuum system equipped with a tungsten hot filament to split H$_2$ into H.[36] Before the initiation of hydrogenation, 100 sccm H$_2$ was introduced into the vacuum chamber and the pressure was stabilized at 10 Torr. The filament current was slowly ramped until the filament power reached 45 W. The VO$_2$ samples were loaded 1.5-2 cm away from the filament for various durations. The sample temperature was controlled at either 25±5 ºC (below the phase transition temperature) or 100±10 ºC (above the phase transition temperature) during the atomic hydrogenation. The schematic of the experimental setup is shown in Figure S1.

**Material characterization.** The Raman spectra were acquired with a Renishaw InVia Raman microscope equipped with a 50× optical objective. A laser with wavelength of 633 nm was employed to characterize the VO$_2$ beams under the constant power of 0.2 mW. The scanning electron microscope (SEM) images were taken with a JEOL 6500F. High-resolution transmission electron microscope (HRTEM) images were obtained on a JEOL 2100F field emission gun TEM.

**Device fabrication and measurement.** Electron beam lithography was employed to define the four electrode leads followed by metal deposition of Ti/Au (3 nm/150 nm) and lift-off. The electrical measurements were conducted in a variable temperature cryostat (Quantum Design PPMS).

*Ab initio* **calculations**. Density Functional Theory, as implemented in the software package CASTEP[37], was used to study hydrogen diffusion in $VO_2$. The Generalized Gradient Approximation (GGA) was used with the exchange-correlation functional of Perdew, Burke, and Ernzerhof (PBE).[38] Ultrasoft pseudo-potentials were used for all atomic species, with a plane-wave basis cutoff of 340 eV. A regular 4×4×4 mesh of k-points was used to sample the Brillouin zone of bulk rutile $VO_2$. The structure of the $VO_2$ lattice was optimized with the Broyden-Fletcher-Goldfarb-Shanno (BFGS) geometry optimization algorithm.[39] The equilibrium positions of hydrogen were found while freezing the positions of V and O ions using constrained BFGS optimization. Path energies were computed by varying the hydrogen position along the diffusion channel and calculating the total energy of the structure. Path parameters were then optimized in order to yield the lowest activation barriers for the diffusion process.

*Acknowledgements*. DN, HJ, and WJH acknowledge support for this work from the US DOE Office of Science/Basic Energy Science; award DE-FG02-06ER46337. AHN acknowledges support from Rober A. Welch Foundation grant C-1818. Additional support for this research was provided by the AFOSR MURI (FA9550-12-1-0035), Sandia National Laboratory (1100745), the Lockheed Martin Corporation through the LANCER IV Program and the ONR MURI program (#00006766, N00014-09-1-1066). The authors acknowledge valuable discussions with Prof. Kevin Kelly and Prof. Jiang Wei.

*Competing financial interests.* The authors declare no competing financial interests. Correspondence and requests for materials should be addressed to natelson@rice.edu.

***Supporting Information Available***: Supporting material includes a schematic of the hydrogenation furnace, transmission electron micrographs of as-grown and hydrogenated beams, additional optical micrographs of nanobeams pre- and post-hydrogenation, and a Raman spectrum of $VO_2$ material after exposure to atomic H at room temperature. This material is available free of charge via the Internet at http://pubs.acs.org.

TOC Figure:

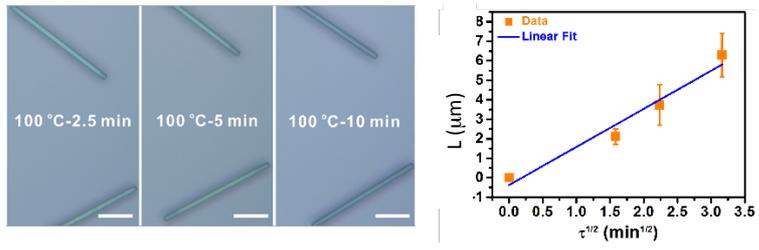

# Hydrogen Diffusion and Stabilization in Single-crystal VO$_2$ Micro/nanobeams by Direct Atomic Hydrogenation – Supporting Information

Jian Lin,[†,‡#] Heng Ji,[§#] Michael W. Swift,[§#] Will Hardy,[§] Zhiwei Peng,[⊥] Xiujun Fan,[‡,⊥] Andriy H. Nevidomskyy[§], James M. Tour[†,‡,⊥] and Douglas Natelson[§,¶*]

[†]*Department of Mechanical Engineering and Material Science;* [‡]*Smalley Institute for Nanoscale Sci. & Tech;* [§]*Department of Physics and Astronomy;* [⊥]*Department of Chemistry;* [¶]*Department of Electrical and Computer Engineering, Rice University, 6100 Main St., Houston, Texas 77005*

# *These people contributed equally to this work.*

*To whom correspondence should be addressed. Emails:* natelson@rice.edu

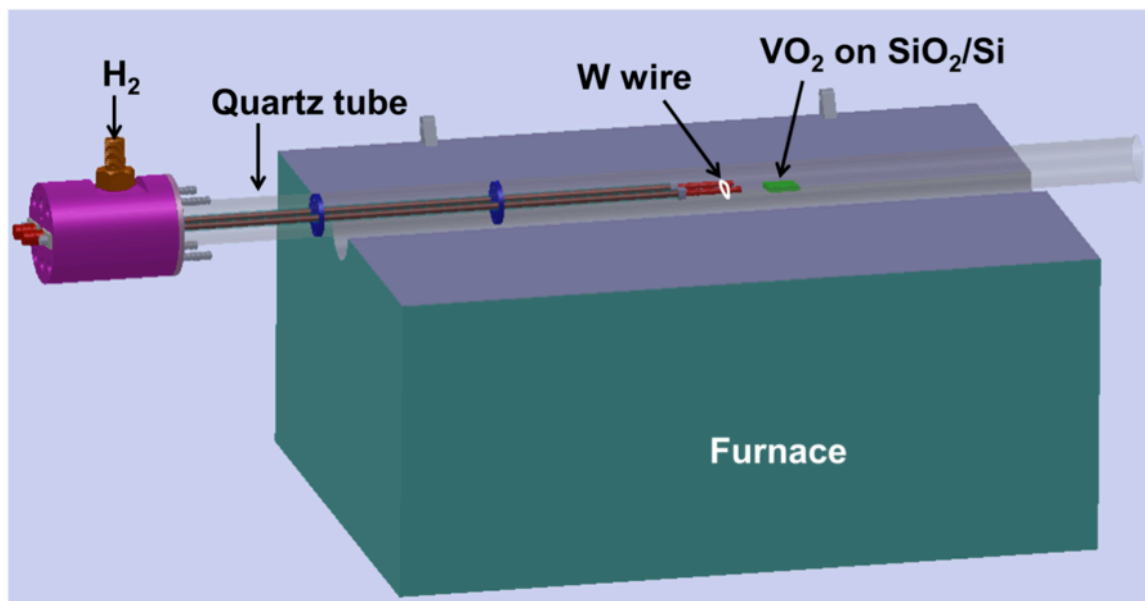

**Figure S1**. Schematic of experimental setup for the hydrogenation of VO$_2$ using atomic hydrogen generated by the hot filament.

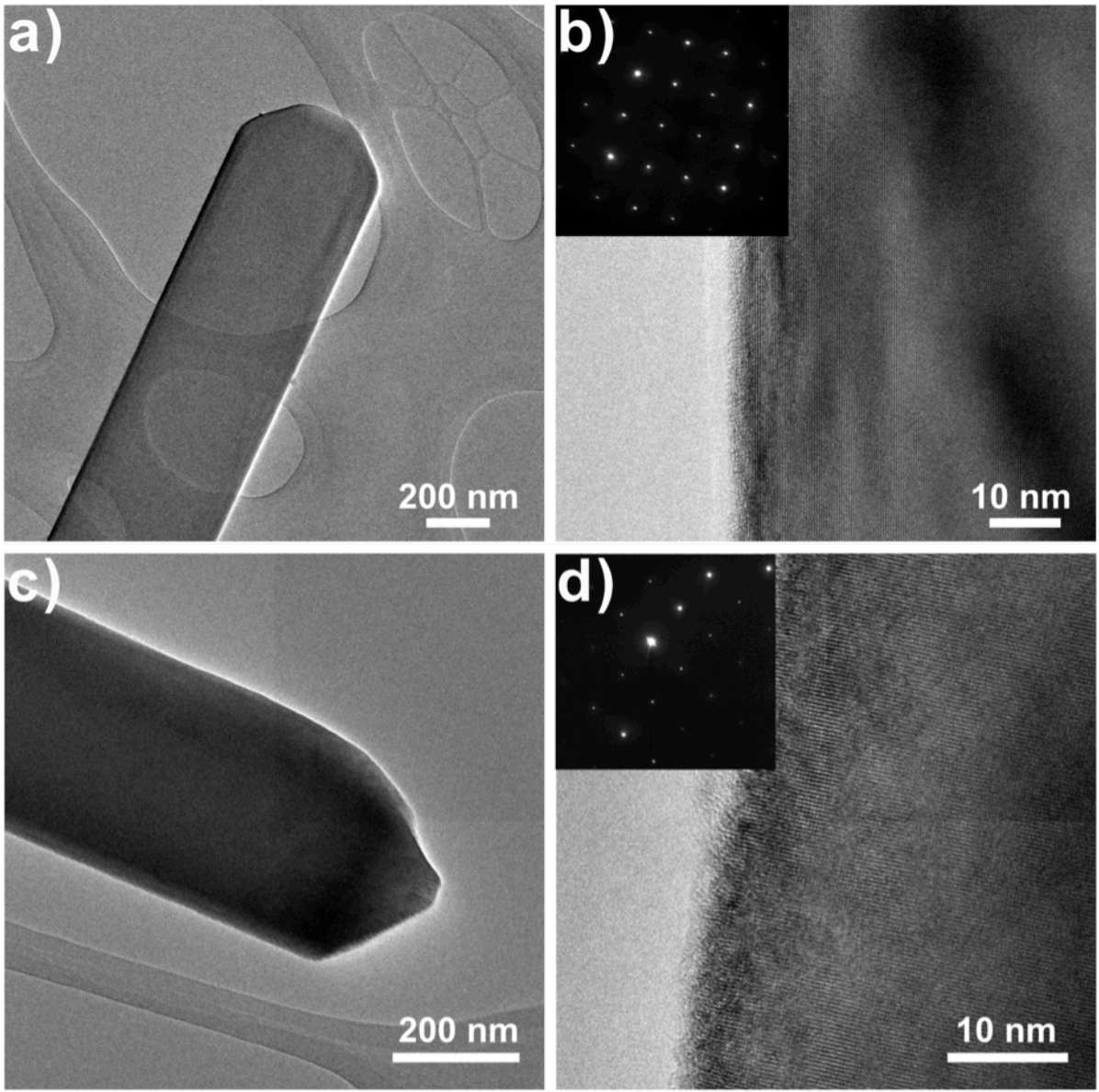

**Figure S2**. (a) Low-magnification TEM image of a pure VO$_2$ nanobeam. (b) High-magnification TEM image of a pure VO$_2$ nanobeam. Inset is the corresponding SAED pattern. (c) Low-magnification TEM image of the hydrogenated VO$_2$ nanobeam. (d) High-magnification TEM image of a hydrogenated VO$_2$ nanobeam. Inset is the corresponding SAED pattern.

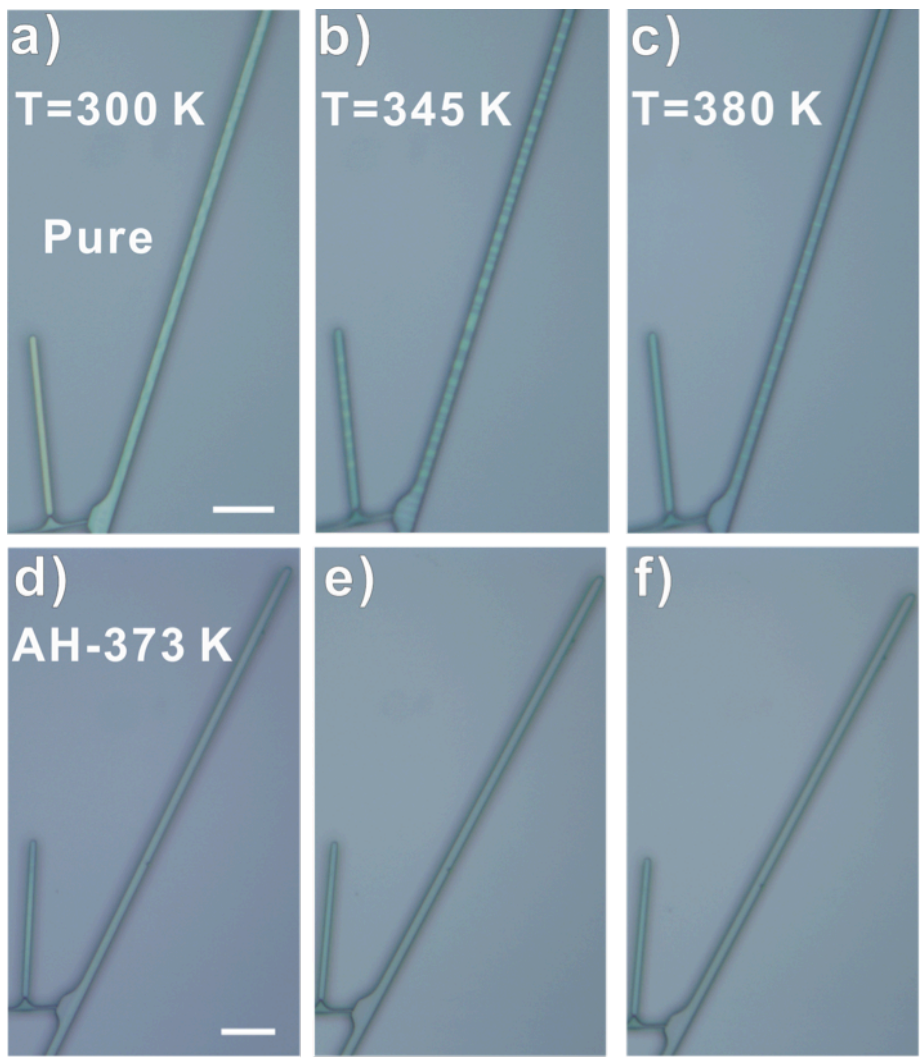

**Figure S3**. Optical images of VO$_2$ beams on Si substrate with 2 μm thermal oxide taken at 300 K, 345 K, 380 K respectively. (a-c) Pure VO$_2$ beams. (d-f) VO$_2$ beams with atomic hydrogenation at 100±10 ºC for 20 min. All of the scale bars are 10 μm.

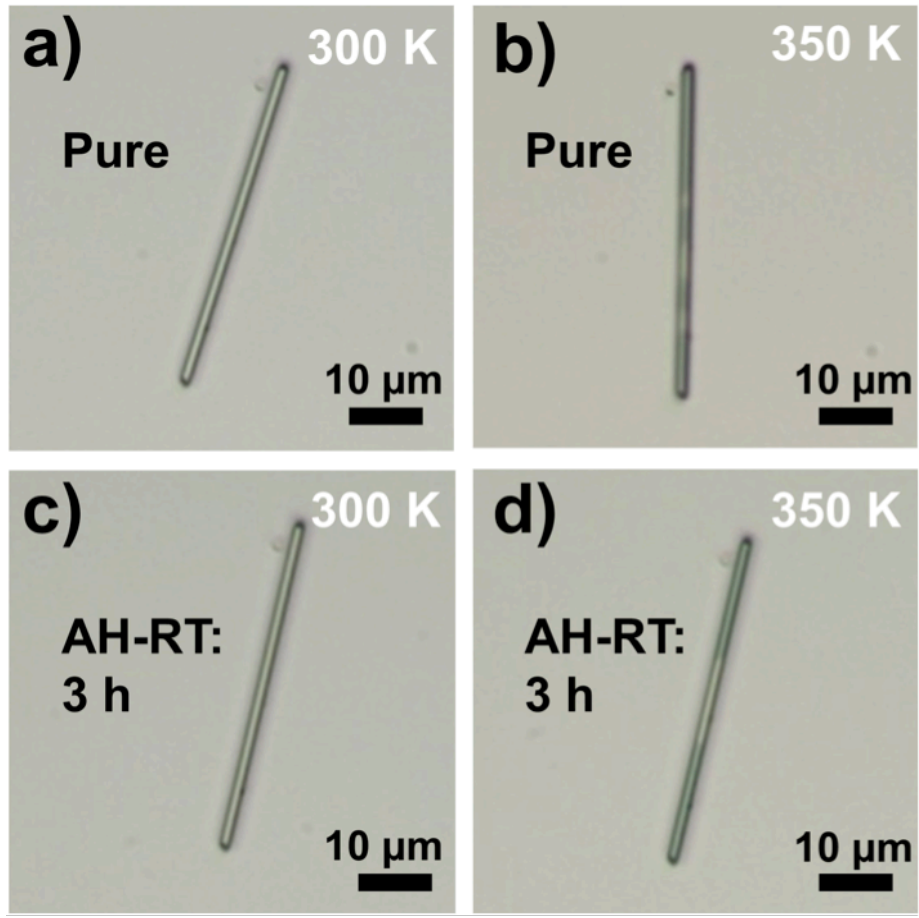

**Figure S4**. a, b) Optical images of pure VO$_2$ microwires taken at 300 K and 350 K (above phase transition temperature). c, d) Optical images of hydrogenated VO$_2$ microwires at room temperature for 3 hours (*e.g.* < 300 K) taken at 300 K and 350 K. They clearly show that hydrogen exposure below the phase transition temperature does not cause metallic phase stabilization.

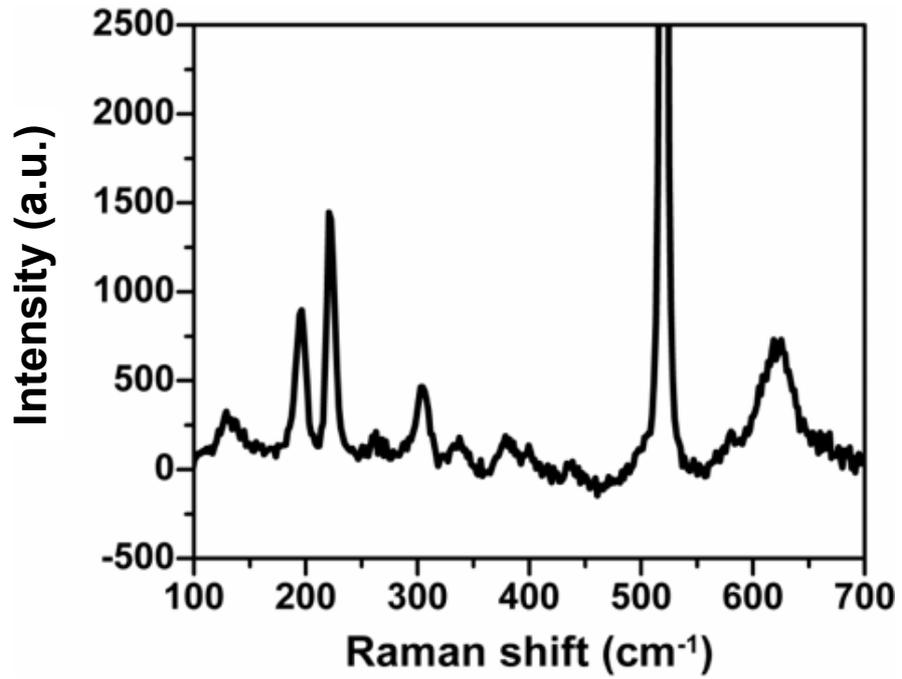

**Figure S5**. Raman spectra of VO$_2$ crystals hydrogenated at room temperature. The characteristics are the same as those seen in spectra of pure VO$_2$ crystals.